\documentclass[twocolumn,preprintnumbers,superscriptaddress,unsortedaddress]{revtex4}
\usepackage{makeidx}
\usepackage{amssymb}
\usepackage{amsmath}
\usepackage{graphicx}
\usepackage{epstopdf}
\usepackage{dcolumn}
\usepackage{bm}
\usepackage{xcolor}
\usepackage{amsfonts}
\usepackage{units}
\usepackage{hyperref}

\begin{document}

\title{Low temperature spectrum of a fiber loop laser}
\author{Eyal Buks}
\email[]{eyal@ee.technion.ac.il}
\affiliation{Andrew and Erna Viterbi Department of Electrical Engineering, Technion,
Haifa 32000 Israel}
\date{\today }

\begin{abstract}
Fiber-based multi-wavelength lasers have a variety of important applications in telecommunication and meteorology. We experimentally study a fiber loop laser with an integrated Erbium doped
fiber (EDF). The output optical spectrum is measured as a function of the
EDF temperature. We find that below a critical temperature of about $10\unit{%
K}$ the measured optical spectrum exhibits a sequence of narrow and
unequally-spaced peaks. An intriguing connection between the peaks' wavelengths and the sequence of prime numbers is discussed. An hypothesis, which attributes the comb formation
to intermode coupling, is explored.
\end{abstract}

\maketitle

\textbf{Introduction} - Erbium doped fibers (EDF) are widely employed in a variety of applications.
Key properties of EDF can be controlled by varying the temperature. The
contribution of Brillouin scattering \cite{Kobyakov_1} to the temperature
dependency has been explored in \cite%
{Le_3611,Anderson_125,Pine_1187,Nikles_1842,Thevenaz_22}.

In this work we study a fiber loop laser with an integrated EDF \cite%
{Antuzevics_1149,haken1985laser}. We measure the emitted optical spectrum as
a function of the EDF temperature \cite{Aubry_2100002,Pizzaia_2352}. Below a
critical temperature of about $10\unit{K}$ the measured optical spectrum
exhibits an unequally-spaced optical comb (USOC) made of a sequence of
narrow peaks. We discuss a possible connection between the observed USOC and
intermode coupling \cite{moloney2018nonlinear}. Theoretical modeling is
employed to explore this connection. The parameters characterizing intermode
coupling are extracted from open loop measurements.

\textbf{Experimental setup} - The experimental setup is schematically depicted in the inset of Fig. \ref%
{FigTempSU}. EDF having length of $20\unit{m}$, absorption of $30$ dB $\unit{%
m}^{-1}$ at $1530\unit{nm}$, and mode field diameter of $6.5\mu \unit{m}$ at 
$1550\unit{nm}$, is cooled down using a cryogen free cryostat. The EDF is
thermally coupled to a calibrated silicon diode serving as a thermometer,
and it is pumped using a $980\unit{nm}$ laser diode (LD) biased with current
denoted by $I_{\mathrm{D}}$. The cold EDF is integrated with a room
temperature fiber loop using a wavelength-division multiplexing (WDM)
device. Two isolators (labeled by arrows in the sketch shown in Fig. \ref%
{FigTempSU}) and a 10:90 output coupler (OC) are integrated in the fiber
loop. The loop frequency $f_{\mathrm{L}}$ (inverse loop period time), which
is measured using a radio frequency spectrum analyzer and a photodetector,
is given by $f_{\mathrm{L}}=c/\left( n_{\mathrm{F}}l_{\mathrm{L}}\right)
=4.963\unit{MHz}$, where $c$ is the speed of light in vacuum, $n_{\mathrm{F}%
}=1.45$ is the fiber refractive index (in the telecom band), and $l_{\mathrm{L}}=41.7\unit{m}$ is
the fiber loop total length. An optical spectrum analyzer (OSA) is connected
to the 10:90 OC.

\begin{figure}[tbp]
\begin{center}
\includegraphics[width=3.2in,keepaspectratio]{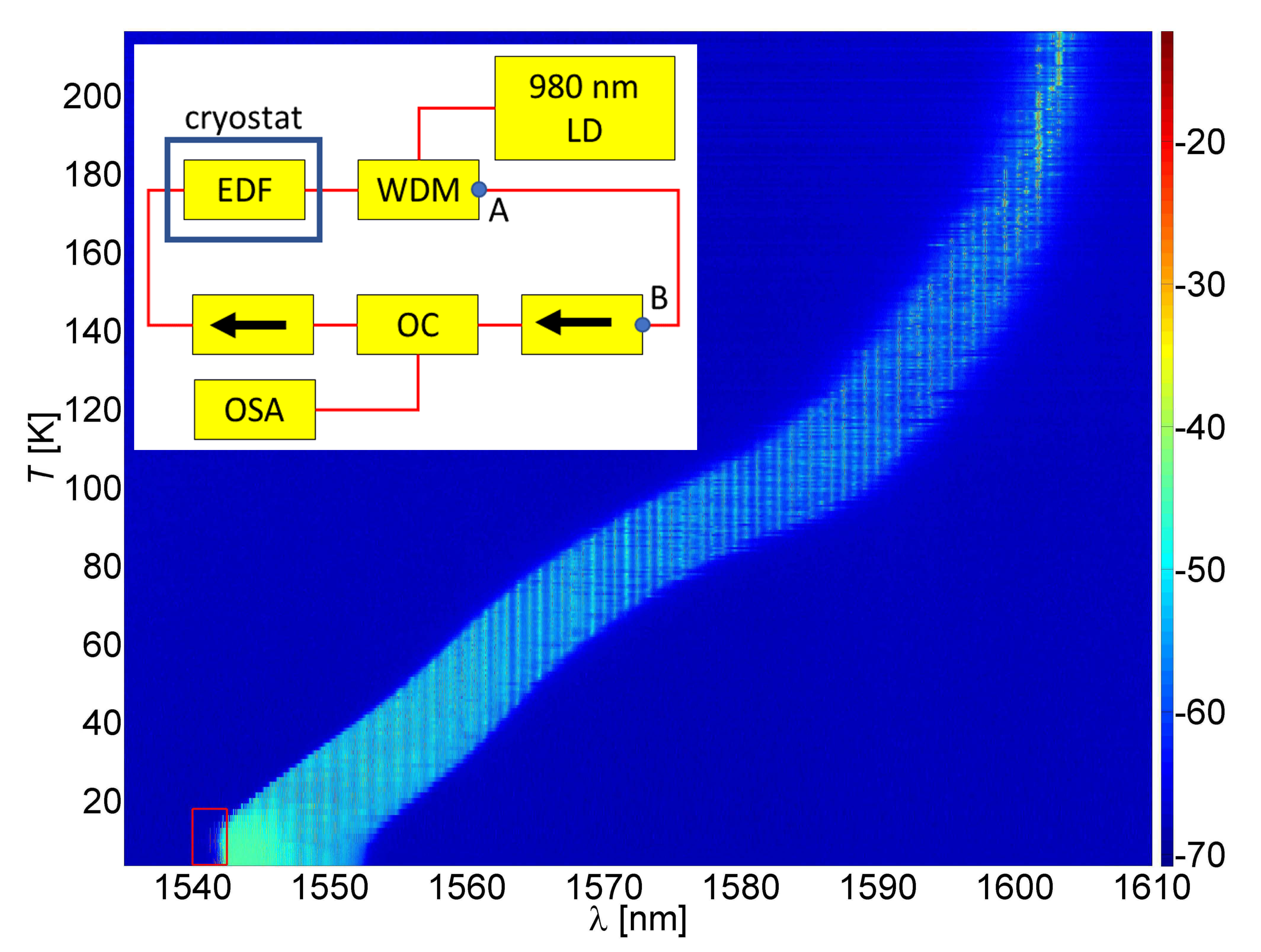}
\end{center}
\caption{{}The measured optical spectrum in dBm units as a function of the
wavelength $\protect\lambda $ and the temperature $T$ with diode current of $%
I_{\mathrm{D}}=200\unit{mA}$. The corresponding diode voltage and diode optical power are $1.38 \unit{V}$ and $ 0.044 \unit{W}$, respectively. The overlaid red rectangle indicates the
region shown in higher resolution in Fig. \protect\ref{FigAPEXvsT_HR}. The
experimental setup is shown in the inset. The EDF is inside the cryostat,
and all other components are at room temperature. The isolator on the right was added to block back-reflected light from the OSA input port.}
\label{FigTempSU}
\end{figure}

\begin{figure*}[tbp]
\begin{center}
\includegraphics[width=7in,keepaspectratio]{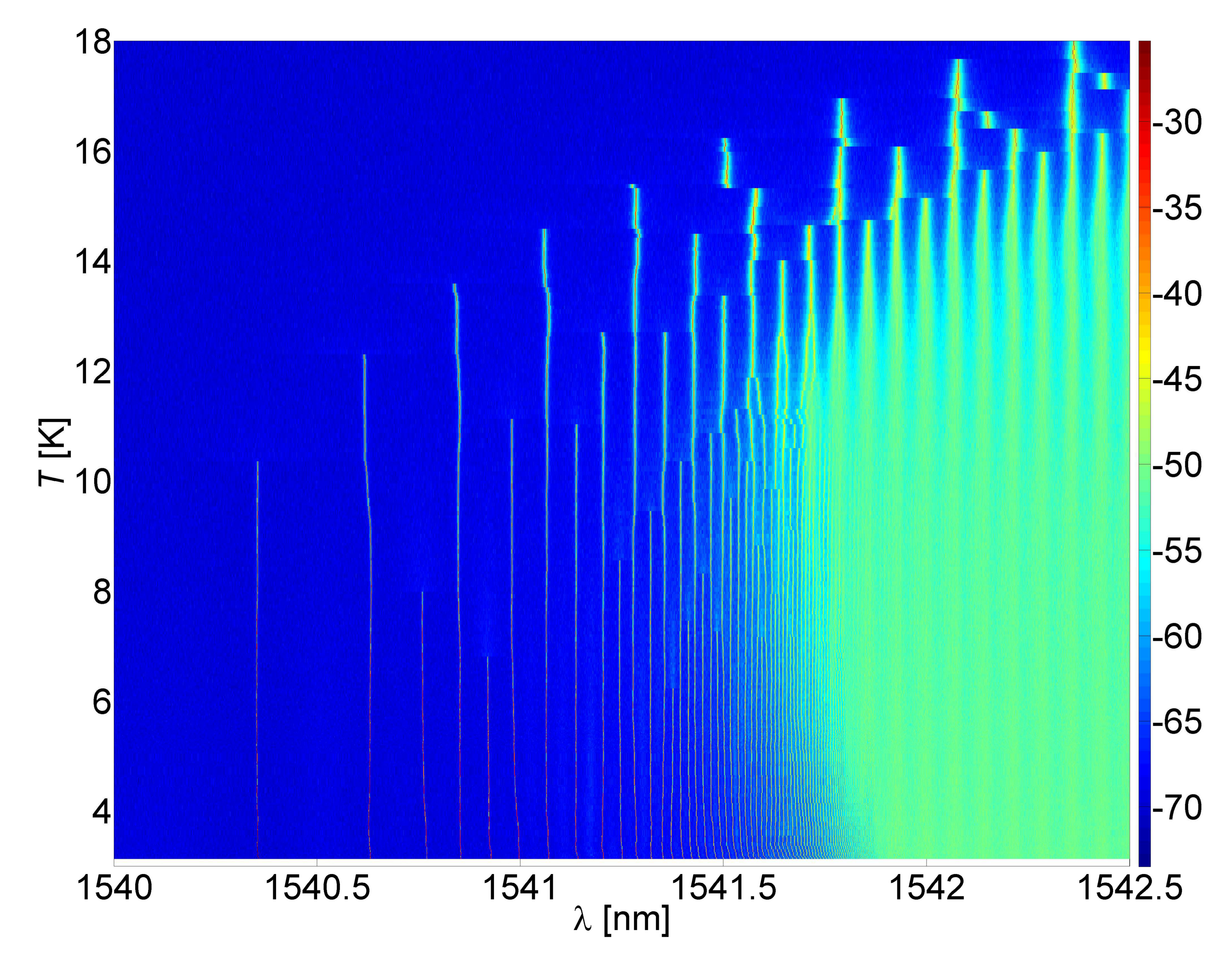}
\end{center}
\caption{{}The optical spectrum in dBm units below $18\unit{K}$. Diode
current for this measurement is $I_{\mathrm{D}}=120\unit{mA}$.}
\label{FigAPEXvsT_HR}
\end{figure*}

\textbf{Temperature dependence} - The measured optical spectrum as a function of the temperature $T$ with
diode current of $I_{\mathrm{D}}=200\unit{mA}$ ($I_{\mathrm{D}}=120\unit{mA}$%
) is shown in Fig. \ref{FigTempSU} (Fig. \ref{FigAPEXvsT_HR}). The low
temperature lasing threshold occurs at $I_{\mathrm{D}}=88\unit{mA}$ [see Fig. \ref{FigAPEXvsID} of the supplementary materials (SM)].

The optical spectrum shown in Fig. \ref{FigTempSU} reveals a transition from
the short to the long EDF limits. The wavelength at which lasing peaks is
denoted by $\lambda _{\mathrm{g}}$. The spectrum shown in Fig. \ref%
{FigTempSU} indicates that $\lambda _{\mathrm{g}}$ decreases as the
temperature is lowered \cite{Kagi_261}, from $1605 \unit{nm}$ at room temperature, to $1540 \unit{nm}$ below $10 \unit{K}$.

It was shown in Ref. \cite%
{Franco_1090} that the wavelength $\lambda _{\mathrm{g}}$ can be determined
by finding the value of $\lambda $ that maximizes the dimensionless variable 
$\eta $, which is given by 
\begin{equation}
\eta =\frac{1-\frac{1}{z_{\mathrm{E}}^{-1}z_{\mathrm{Fe}}}}{1+\frac{z_{%
\mathrm{A}}^{-1}}{z_{\mathrm{E}}^{-1}}}\;,  \label{eta EDFA T}
\end{equation}%
where $z_{\mathrm{E}}^{-1}$ ($z_{\mathrm{A}}^{-1}$) is the emission
(absorption) inverse length at wavelength $\lambda $, and $z_{\mathrm{Fe}%
}=z_{\mathrm{F}}/\log \gamma _{\mathrm{L}}$ represents an effective value
for the EDF length $z_{\mathrm{F}}$, where $\gamma _{\mathrm{L}}\geq 1$ is
the loop loss coefficient, which is mainly determined by the output coupler
splitting ratio. In thermal equilibrium the ratio $z_{\mathrm{A}}^{-1}/z_{%
\mathrm{E}}^{-1}$ at wavelength $\lambda $ is given by the Einstein
(McCumber) relation $z_{\mathrm{A}}^{-1}/z_{\mathrm{E}}^{-1}=e^{-\rho }$,
where $\rho =\lambda _{\mathrm{T}}\left( \lambda _{0}^{-1}-\lambda
^{-1}\right) $, $\lambda _{0}$ is the wavelength for which $z_{\mathrm{A}%
}^{-1}=z_{\mathrm{E}}^{-1}$, $\lambda _{\mathrm{T}}=hc/\left( n_{\mathrm{F}%
}k_{\mathrm{B}}T\right) $\ is the thermal wavelength, where $h$ is the
Planck's constant, $k_{\mathrm{B}}$ is the Boltzmann's constant, and $T$ is
the temperature.

The relative importance of the term $1/\left( 1+z_{\mathrm{A}}^{-1}/z_{%
\mathrm{E}}^{-1}\right) \equiv f$ in Eq. (\ref{eta EDFA T}) depends on the
ratio $z_{\mathrm{E}}^{-1}z_{\mathrm{Fe}}$ between the effective EDF length $%
z_{\mathrm{Fe}}$ and the emission length $z_{\mathrm{E}}$. In the long EDF
limit (i.e. when $z_{\mathrm{E}}^{-1}z_{\mathrm{Fe}}\gg 1$)\ the term $f$
has a relatively large influence on the lasing wavelength $\lambda _{\mathrm{%
g}}$. In our experiment, the long EDF limit corresponds to temperatures
above $200\unit{K}$, for which $\lambda _{\mathrm{g}}\simeq 1605\unit{nm}$
(see Fig. \ref{FigTempSU}). At this value of $\lambda _{\mathrm{g}}$
absorption is strongly suppressed, i.e. $z_{\mathrm{A}}^{-1}\ll z_{\mathrm{E}%
}^{-1}$, and consequently a relatively large value for $f$ is obtained. In
the opposite limit of low temperatures below $10\unit{K}$, the ratio $z_{%
\mathrm{E}}^{-1}z_{\mathrm{Fe}}$ decreases, and consequently the lasing band
is shifted to the region where emission cross section peaks near $\lambda _{%
\mathrm{g}}\simeq 1540\unit{nm}$ (see Fig. \ref{FigTempSU}) \cite%
{Desurvire_547,Desurvire_246,Zyskind_869}.

\begin{figure}[tbp]
\begin{center}
\includegraphics[width=3.2in,keepaspectratio]{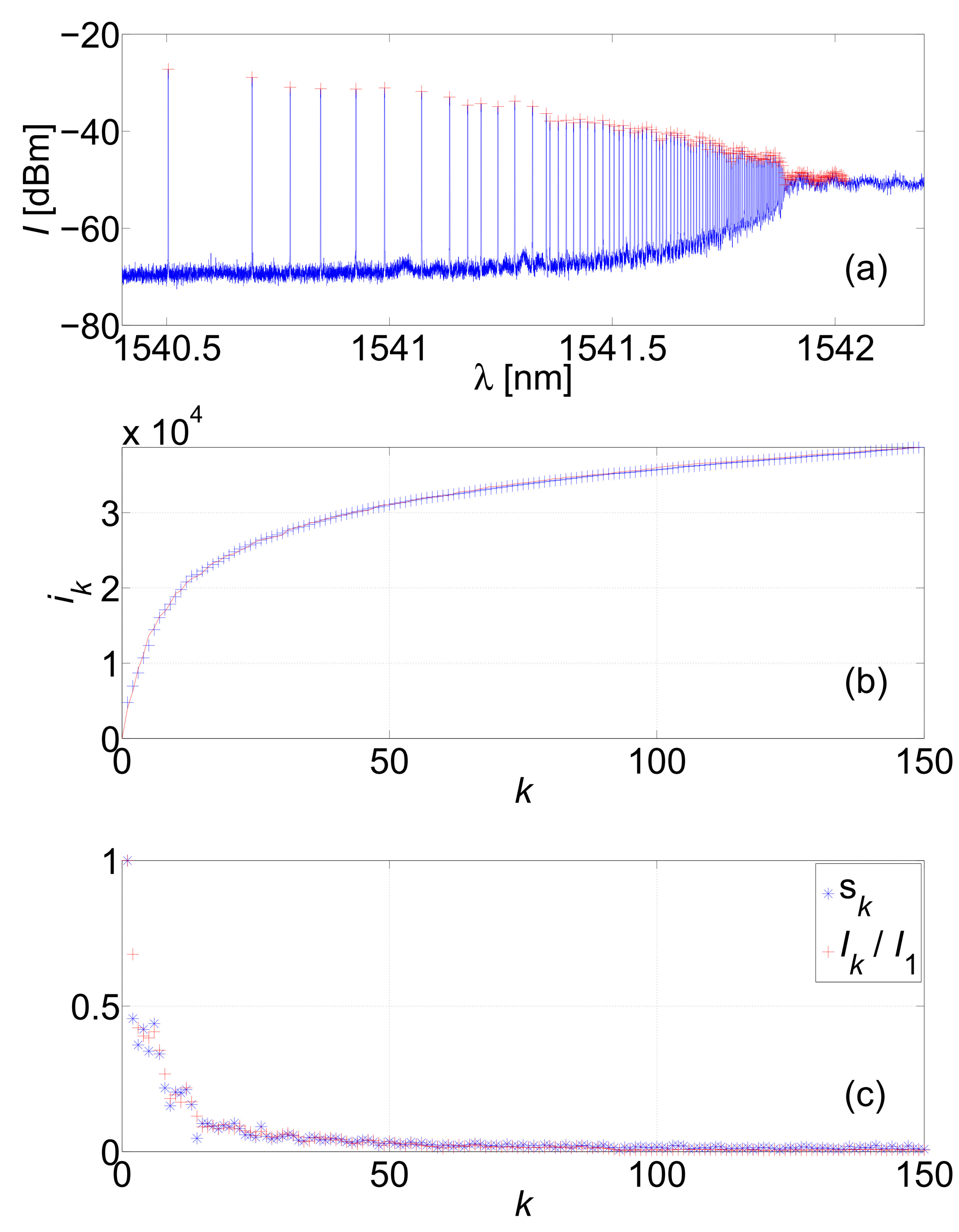}
\end{center}
\caption{{}The USOC sequence. The LD is biased with a current of $I_{\mathrm{%
D}}=120\unit{mA}$. (a) The optical spectrum (measured by the OSA) at the
base temperature of $3\unit{K}$. (b) Comparison between the measured values
of the dimensionless sequence $i_{k}=\left( f_{0}-f_{k}\right) /f_{\mathrm{L}%
}$ (crosses), and the values of $n_{k}$ calculated using Eq. (\protect\ref%
{i_n}) of the SM (solid line). (c) The normalized wavelength gaps $%
s_{k}=\left( \protect\lambda _{k}-\protect\lambda _{k-1}\right) /\left( 
\protect\lambda _{1}-\protect\lambda _{0}\right) $ (stars) and the USOC
measured normalized intensities $I_{k}/I_{1}$ (crosses) as a function of $k$%
. }
\label{FigComb_in_In}
\end{figure}

Near wavelength of $1540\unit{nm}$ and below a critical temperature of about 
$10\unit{K}$ the measured optical spectrum exhibits narrow peaks at a
sequence of wavelengths denoted by $\left\{ \lambda _{k}\right\} $, where $%
k=0,1,2,\cdots $ [see Fig. \ref{FigAPEXvsT_HR} and Fig. \ref{FigComb_in_In}%
(a)]. For the data presented in Fig. \ref{FigComb_in_In}(a) $\lambda
_{0}=1540.5039\unit{nm}$ and $\lambda _{1}-\lambda _{0}=0.1875\unit{nm}$.
The frequency $f_{k}$ associated with wavelength $\lambda _{k}$ is given by $%
f_{k}=c/\lambda _{k}$. For the smallest wavelength spacings that can be
reliably resolved $\lambda _{k+1}-\lambda _{k}\simeq 1.7\unit{pm}$ and $%
f_{k+1}-f_{k}\simeq -200\unit{MHz}$, with $k\simeq 150$. The frequency
associated with the gap between $\lambda _{0}$ and $\lambda _{1}$ is given
by $f_{1}-f_{0}=23.69\unit{GHz}$.

A plot of the dimensionless sequence $i_{k}=\left( f_{0}-f_{k}\right) /f_{%
\mathrm{L}}$ is shown in Fig. \ref{FigComb_in_In}(b). This sequence is further explored in the SM. In particular, Eq. (S14) of the SM reveals an intriguing connection between $i_{k}$ and the sequence of prime numbers.

The intensity of the
USOC peak occurring at wavelength $\lambda _{k}$ is denoted by $I_{k}$. The
plot in Fig. \ref{FigComb_in_In}(c) compares the USOC measured normalized
wavelength gaps $s_{k}=\left( \lambda _{k}-\lambda _{k-1}\right) /\left(
\lambda _{1}-\lambda _{0}\right) $ and the USOC measured normalized
intensities $I_{k}/I_{1}$. The comparison indicates that to a good
approximation $s_{k}=I_{k}/I_{1}$, i.e. the ratio $\left( \lambda
_{k}-\lambda _{k-1}\right) /I_{k}$ is nearly a constant.

No change in the measured optical spectrum is detected when a magnetic field up to $0.15 \unit{T}$ is externally applied. Note that strong
effect of magnetic field on the optical decoherence rate in an EDF was found
at low temperatures using the method of two-pulse photon echoes \cite%
{Macfarlane_033602,Veissier_195138}.

\textbf{Open loop} - Open loop measurements are performed by disconnecting the fiber between the
points labeled as 'A' and 'B' in the inset of Fig. \ref{FigTempSU},
connecting OSA to 'A' and an optical source to 'B'. As is discussed below,
open loop measurements allow the characterization of EDF gain and intermode
coupling.

\begin{figure}[tbp]
\begin{center}
\includegraphics[width=3.2in,keepaspectratio]{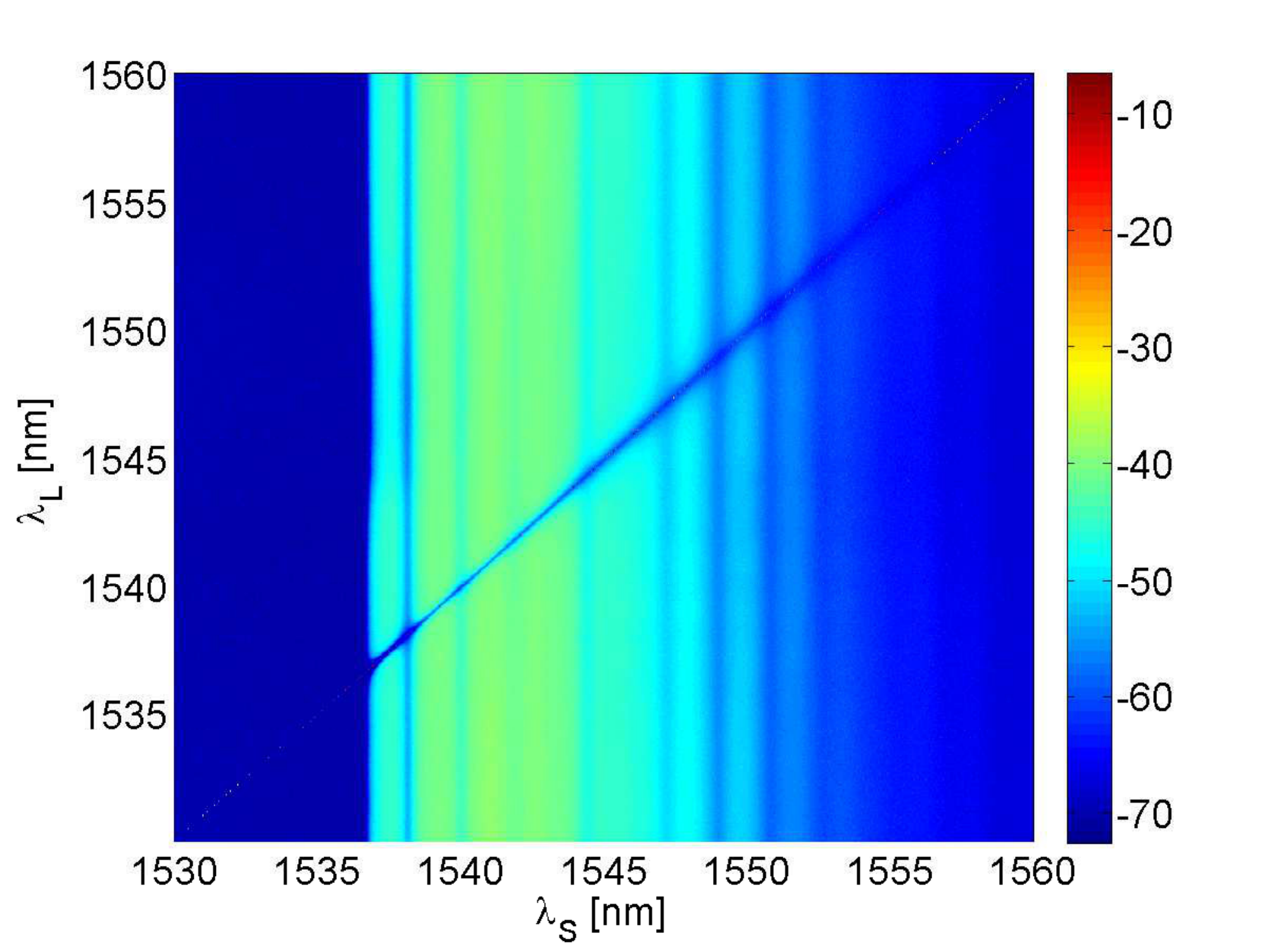}
\end{center}
\caption{{}Open loop gain. The transmitted light optical spectrum is shown
in dBm units as a function of the tunable laser wavelength $\protect\lambda _{%
\mathrm{L}}$. Tunable laser linewidth is $1.6 \unit{pm}$, its power is $-4.1$ dBm, the temperature is $T=2.9\unit{K%
}$, diode current is $I_{\mathrm{D}}=150\unit{mA}$, and its optical linewidth is about $1 \unit{nm}$.}
\label{FigLamSLamL}
\end{figure}

For the measurements shown in Fig. \ref{FigLamSLamL} the source connected to
'B' is a narrow band laser having a tunable wavelength $\lambda _{\mathrm{L}%
} $. The measured transmitted light optical spectrum is shown in Fig. \ref%
{FigLamSLamL} as a function of $\lambda _{\mathrm{L}}$. The signal
suppression near $\lambda _{\mathrm{L}}$ is attributed to the effects of
gain saturation and hole burning \cite{Rittner_1}. A similar measurement performed with a closed loop is presented by Fig. \ref{FigHB} of the SM.

\begin{figure}[tbp]
\begin{center}
\includegraphics[width=3.2in,keepaspectratio]{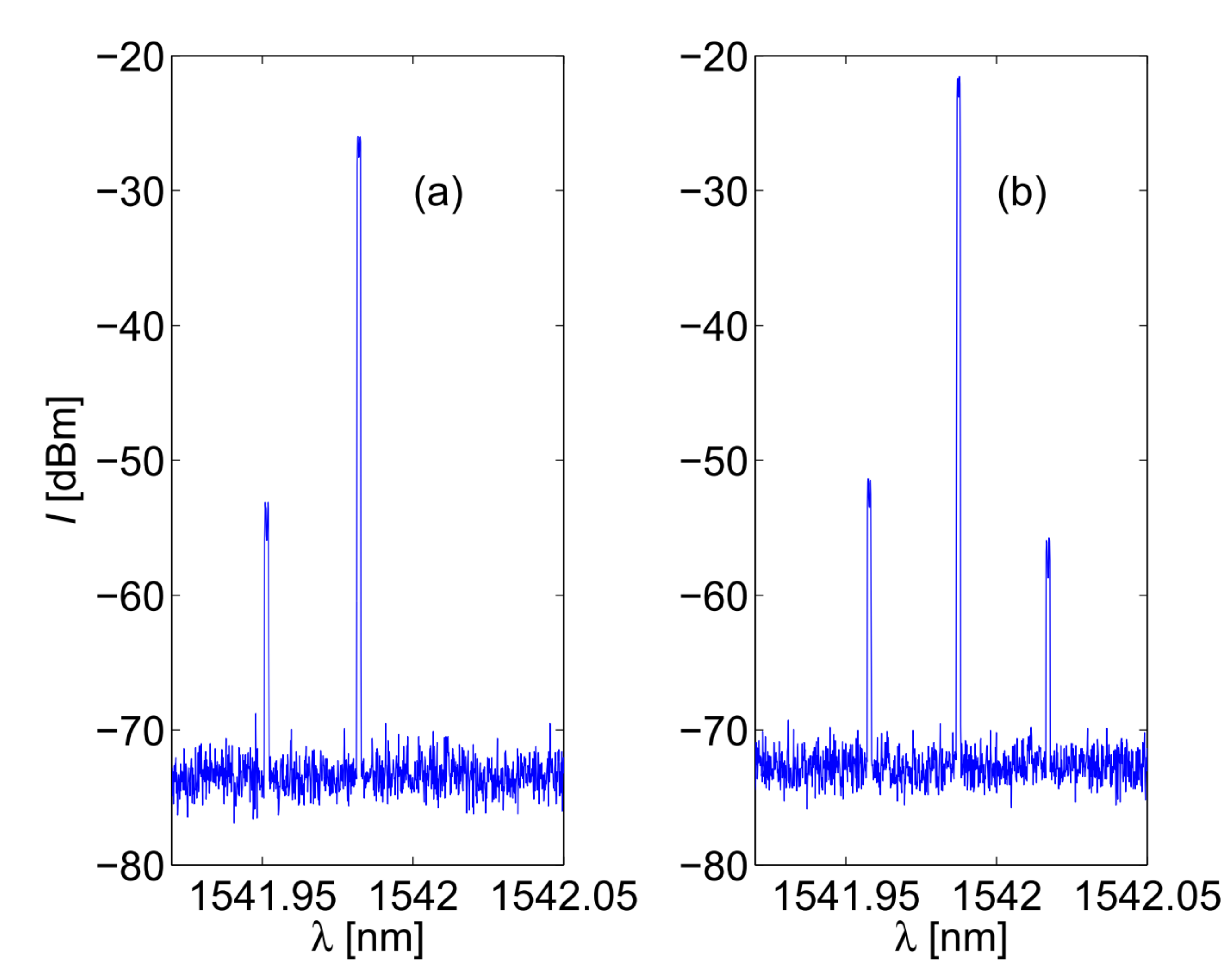}
\end{center}
\caption{{}Open loop IMD. The transmitted light optical spectrum (a) below
threshold $I_{\mathrm{D}}=94\unit{mA}$, and (b) above threshold $I_{\mathrm{D%
}}=144\unit{mA}$. The laser power is $14.9$ dBm, the laser wavelength is $%
1541.982\unit{nm}$, the temperature is $T=2.9\unit{K}$, the FSR modulator
driving frequency is $3.792\unit{GHz}$, and the corresponding optical
sideband wavelength detuning is $\protect\lambda _{\mathrm{d}}=30\unit{pm}$.}
\label{FigOL}
\end{figure}

Intermode coupling can be explored using the method of intermodulation
(IMD). This is done by injecting two monochromatic tones into the system
under study. The first one, which has a relatively large amplitude, and a
wavelength denoted by $\lambda _{\mathrm{p}}$, is commonly refereed to as
the \textit{pump}. The second one is a relatively low-amplitude \textit{%
signal} tone having wavelength $\lambda _{\mathrm{s}}=\lambda _{\mathrm{p}%
}-\lambda _{\mathrm{d}}$, where $\lambda _{\mathrm{d}}$ is the detuning
wavelength, which is assumed to be small $\left\vert \lambda _{\mathrm{d}%
}\right\vert \ll \lambda _{\mathrm{p}}$. Nonlinear frequency mixing can be
characterized by measuring the response at the \textit{idler} wavelength $%
\lambda _{\mathrm{i}}=\lambda _{\mathrm{p}}+\lambda _{\mathrm{d}}$. For the
IMD measurements presented in Fig. \ref{FigOL}, an optical modulator based on
a ferrimagnetic sphere resonator (FSR) is employed \cite{Nayak_193905}.
This device can generate single sideband modulation provided that the input
polarization (of the laser light injected into the FSR) is properly tuned
[see Fig. \ref{FigOL}(a)] \cite{Nayak_193905}. Frequency mixing between
the pump and signal input tones occurring in the EDF gives rise to an idler tone at the output.
The idler peak can be detected in the transmitted light optical spectrum
when the diode current $I_{\mathrm{D}}$ is tuned above its threshold value
[see Fig. \ref{FigOL}(b)]. The measured idler intensity allows the extraction
of IMD gain of the medium, which, in turn, determines the intermode coupling
rates.

\textbf{Equations of motion} - Intermode coupling is theoretically explored by deriving equations of
motion. The (assumed slowly varying) complex amplitude of the fiber loop $m$'th mode is denoted by $%
c_{m}$. Consider the case where the time evolution of $c_{m}$ is governed by 
\cite{Haus_1173}%
\begin{equation}
\dot{c}_{m}=\Gamma _{\mathrm{d}}m^{2}c_{m}-\Gamma _{\mathrm{c}%
}V_{m}+g_{m}c_{m}\;,  \label{a_m dot}
\end{equation}%
where overdot denotes a time derivative, both the dispersion rate $\Gamma _{%
\mathrm{d}}=\Gamma _{\mathrm{d}}^{\prime }+i\Gamma _{\mathrm{d}}^{\prime
\prime }$ and coupling rate $\Gamma _{\mathrm{c}}=\Gamma _{\mathrm{c}%
}^{\prime }+i\Gamma _{\mathrm{c}}^{\prime \prime }$ are complex ($\Gamma _{%
\mathrm{d}}^{\prime }$ , $\Gamma _{\mathrm{d}}^{\prime \prime }$, $\Gamma _{%
\mathrm{c}}^{\prime }$ and $\Gamma _{\mathrm{c}}^{\prime \prime }$ are all
real), and where the interaction term $V_{m}$ is given by%
\begin{equation}
V_{m}=\sum_{n^{\prime }-n^{\prime \prime }+n^{\prime \prime \prime
}=m}c_{n^{\prime }}^{{}}c_{n^{\prime \prime }}^{\ast }c_{n^{\prime \prime
\prime }}^{{}}\;.
\end{equation}%
Due to strong temperature dependency of line-widths of optical transitions in the EDF  \cite{Chu_966}, multimode lasing becomes possible at low temperatures \cite{perez2013multi,haken1985laser}. This is taken into account by allowing the gain $g_{m}$ in the master equation (\ref{a_m dot}), which is commonly assumed to be mode-independent, to vary with $m$. The assumed dependency is given by $g_{m}=\left(
1+g_{0}\right) /\left( 1+H_{m}/I_{\mathrm{sat}}\right) -1$, where $g_{0}$ is
the low intensity gain, and $I_{\mathrm{sat}}$ is the saturation intensity.
The open loop gain measurements [see Fig. \ref{FigLamSLamL}] can be well
fitted to $g_{m}$ when the effective $m$'th intensity $H_{m}$\ is taken to
be given by $H_{m}=\sum_{n}h_{n,m}\left\vert c_{n}\right\vert ^{2}$, where $%
h_{n,m}=\delta _{n,m}+\alpha _{\mathrm{H}}\left( 1-\delta _{n,m}\right)
\left\vert n-m\right\vert ^{-1}$, and $\alpha _{\mathrm{H}}$ is a positive
constant. For this effective intensity $H_{m}$, the hole that is burned by
a single excited mode having amplitude $c_{n}$ contains about $\alpha _{%
\mathrm{H}}I_{\mathrm{sat}}^{-1}\left\vert c_{n}\right\vert ^{2}$ modes. The
value of $\Gamma _{\mathrm{c}}$ is determined from IMD measurements [see Fig. \ref{FigOL}].

\begin{figure}[ht]
\begin{center}
\includegraphics[width=3.2in,keepaspectratio]{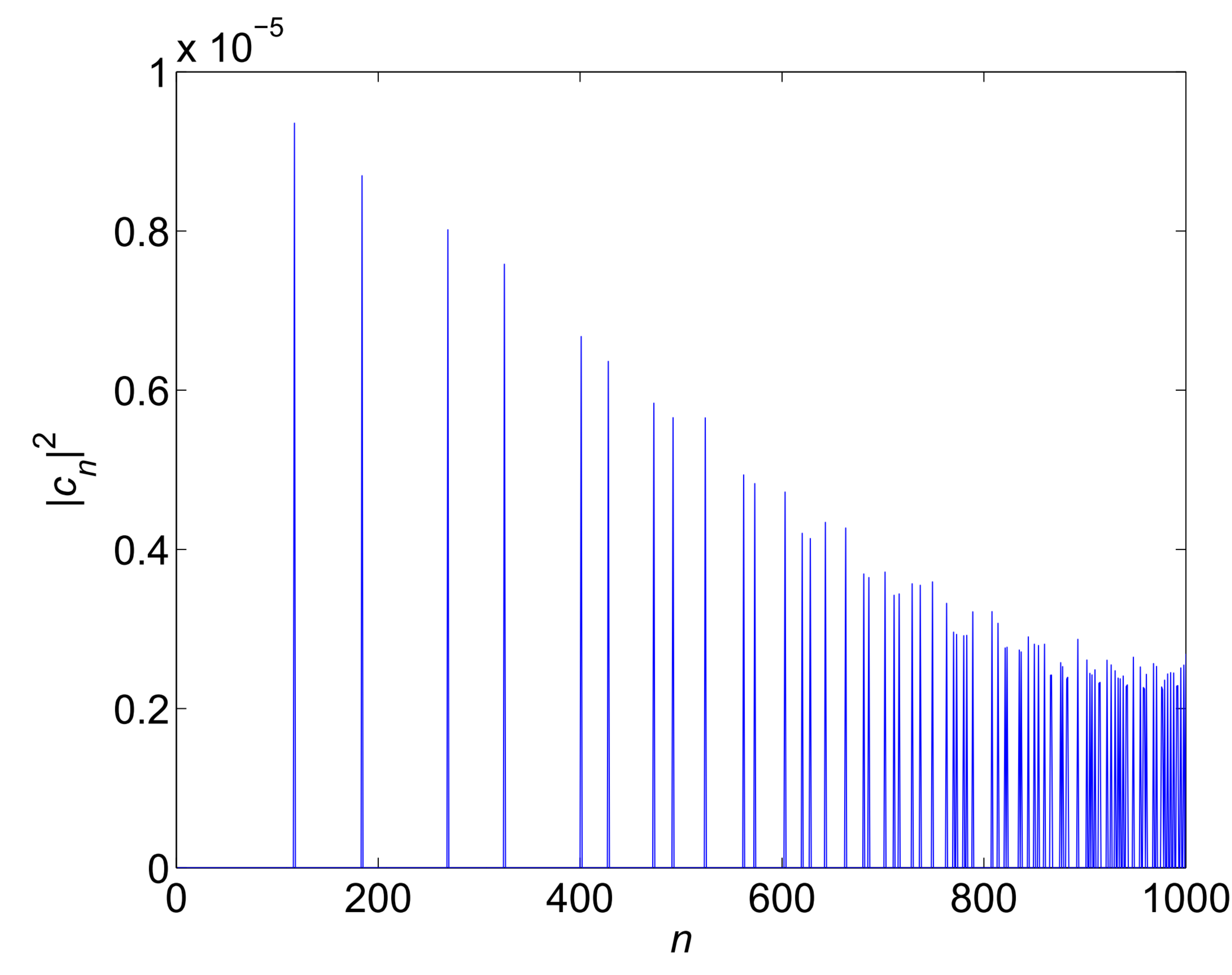}
\end{center}
\caption{Numerical solution of the set of coupled equations (\protect\ref%
{a_m dot}). The mode index number is denoted by $n$. For this example the total number of modes is $1000$, $\Gamma _{%
\mathrm{d}}=10^{-7}\times \left( 1+0.1i\right) $, $\Gamma _{\mathrm{c}}=0.05$%
, $g_{0}=1.2$, $I_{\mathrm{sat}}=10^{-5}$ and $\protect\alpha _{\mathrm{H}%
}=200$.}
\label{FigSatG}
\end{figure}

A numerical solution example for the coupled equations (\ref{a_m dot}) is shown in Fig. \ref{FigSatG}. For this example, the gap between peaks (in units of spacing between neighboring modes) varies from about $70$ (near $n=0$) to about $2$ (near $n=1000$), where $n$ denotes the mode index number. This example demonstrates that
this simple model can account for a spontaneous generation of a USOC. However, further study is needed to explore the temperature dependency of the model's parameters, in order to account for the experimental observation that the USOC becomes visible only below the critical temperature of about $10\unit{K}$.

\textbf{Summary} - In summary, an USOC is observed when the EDF is cooled down below $10\unit{K}
$. An unequally spaced wavelength sequence can be generated by a variety of
mechanisms, including Brillouin scattering, gain and loss grating (due to
back reflection at the splicing points at both ends of the EDF), and lasing
without population inversion. However, for all these cases, the
theoretically predicted sequences were found to be inconsistent with the
experimental results. Open loop measurements and theoretical results are
presented to support the hypothesis that intermode coupling is the
underlying mechanism responsible for the USOC formation. An intriguing connection between $\left\{ \lambda _{k}\right\} $ and the sequence of prime numbers is discussed in the SM. Future work will be devoted to explore the interplay between USOC formation, hole burning and external injection. The extremely high
stability of the USOC can be exploited for some novel applications.

This work was supported by the Israeli science foundation. Preliminary
measurements of a device similar to the one under study in this paper have
been performed by A. Becker together with EB (these measurements are not
included in this paper). The data that support the findings of this study are available from the corresponding author upon reasonable request.

\bibliographystyle{IEEEtran}
\bibliography{Eyal_Bib}

%%%%%%%%%% Merge with supplemental materials %%%%%%%%%%
%\pagebreak
\widetext \widetext \clearpage

\begin{center}
\textbf{\large Supplemental Materials: Low temperature spectrum of a fiber loop laser}

{\small {Eyal Buks} }

{\small {Andrew and Erna Viterbi Department of Electrical Engineering,
Technion, Haifa 32000 Israel} }
\end{center}

%%%%%%%%%% Merge with supplemental materials %%%%%%%%%%
%%%%%%%%%% Prefix a "S" to all equations, figures, tables and reset the counter %%%%%%%%%%
\setcounter{equation}{0}
\setcounter{figure}{0}
\setcounter{table}{0}
\setcounter{section}{0}
\setcounter{page}{1}
\makeatletter \renewcommand{\theequation}{S\arabic{equation}} \renewcommand{\thefigure}{S\arabic{figure}}
\renewcommand{\thesection}{S\arabic{section}}
\renewcommand{\bibnumfmt}[1]{[S#1]}
\renewcommand{\citenumfont}[1]{S#1} 
%%%%%%%%%% Prefix a "S" to all equations, figures, tables and reset the counter %%%%%%%%%%

The unequally-spaced optical comb (USOC) is measured as a function of the diode current $I_{\mathrm{D}}$ in section \ref{SecAPEXvsID}. The effect of laser injection into the loop  is discussed in section \ref{SecLI}. To study the USOC wavelength selection process, the short-time scale dynamics of the system are theoretically explored in section \ref{SecSTSD}. The measured pattern of the USOC wavelength sequence $\left\{ \lambda _{k}\right\} $ is discussed in sections \ref{SecSequence} and \ref{SecRUSU}. We find that the observed pattern $\left\{ \lambda _{k}\right\} $ can be attributed to a process, in which the intermode coupling contribution
to the system's total energy is minimized, under the constrain that the total optical intensity is given. An intriguing connection between $\left\{ \lambda _{k}\right\} $ and the sequence of prime numbers is discussed.

\section{Diode current}
\label{SecAPEXvsID}

\begin{figure*}[b]
	\begin{center}
		\includegraphics[width=6.8in,keepaspectratio]{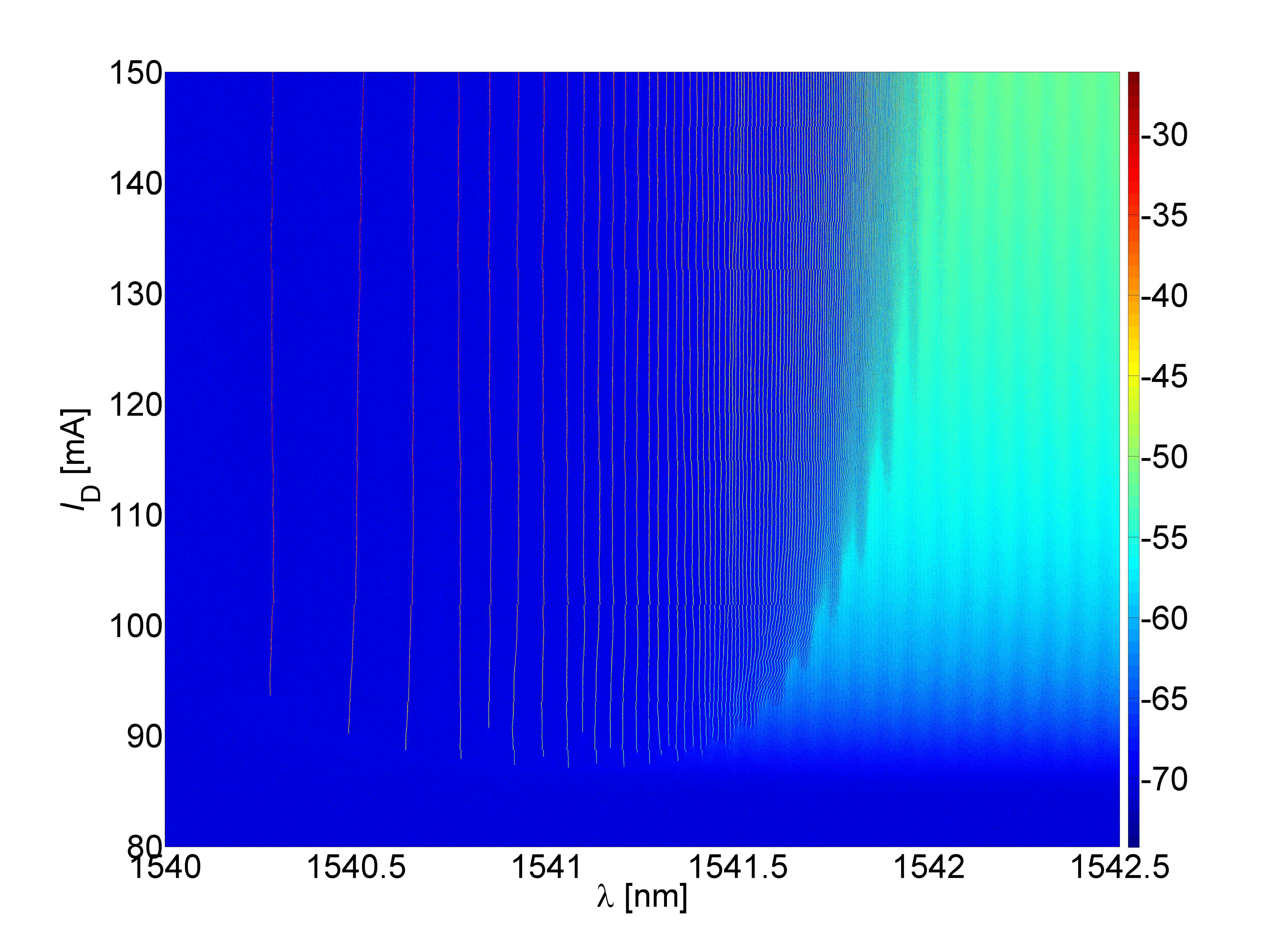}
	\end{center}
	\caption{{}The optical spectrum in dBm units as a function of  diode current $I_{\mathrm{D}}$ at temperature of $2.9 \unit{K}$.}
	\label{FigAPEXvsID}
\end{figure*}

The measured optical spectrum is shown in Fig. \ref{FigAPEXvsID} as a function of the diode current $I_{\mathrm{D}}$. The lasing threshold occurs at $I_{\mathrm{D}}=88\unit{mA}$.

\section{Laser injection}
\label{SecLI}

While the laser injection measurements that are described in the main text are performed with an open loop (see Fig. \ref{FigLamSLamL} of the main text), the plot in Fig. \ref{FigHB} below displays the effect of laser injection into the closed loop. USOC suppression near the laser wavelength of $\protect\lambda_{\mathrm{L}}=1541.5\unit{nm}$ is attributed to the effect of gain saturation.

\begin{figure*}[b]
	\begin{center}
		\includegraphics[width=6.8in,keepaspectratio]{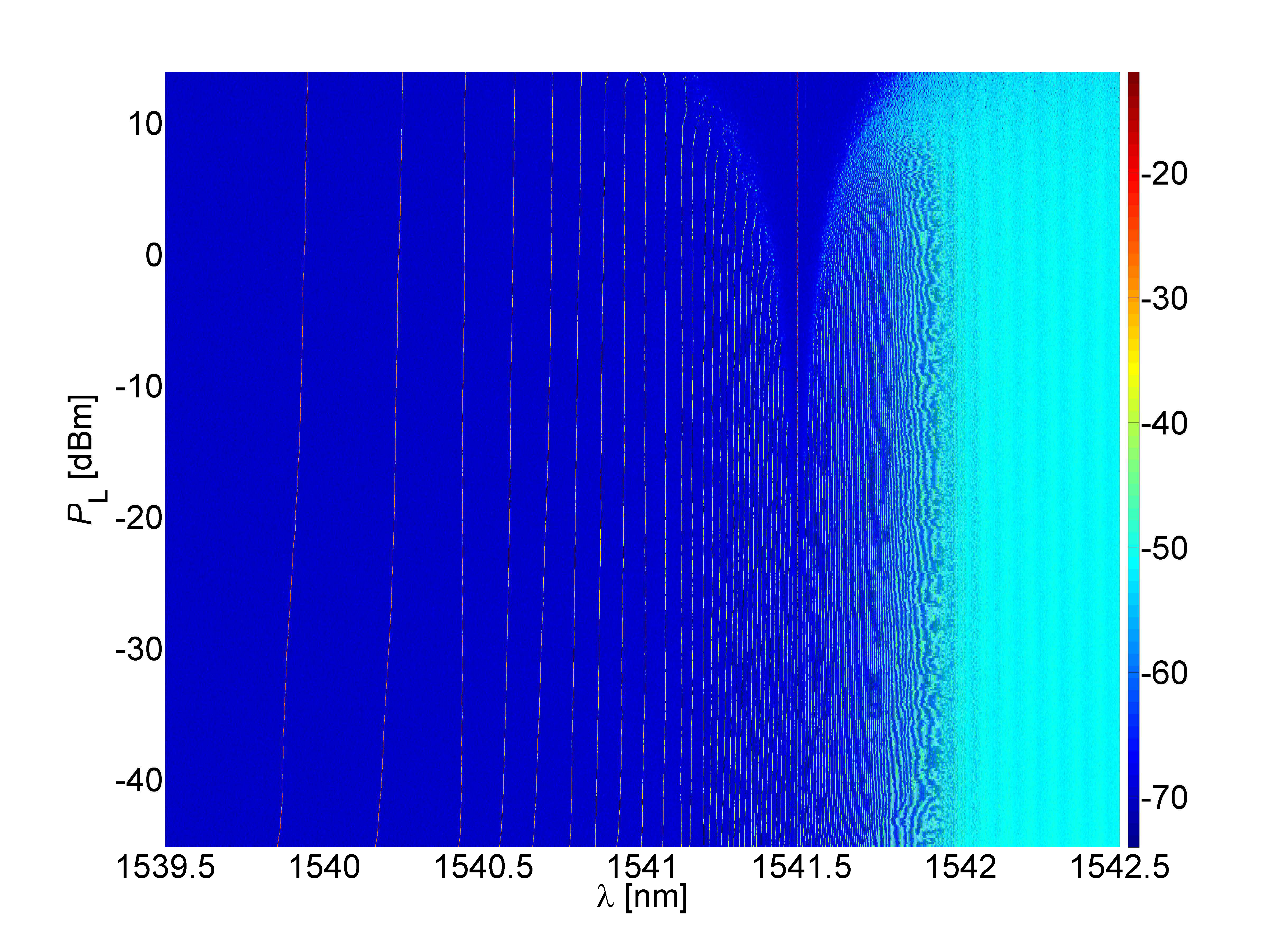}
	\end{center}
	\caption{{}Laser injection with a closed loop. The laser, having power $P_{\mathrm{L}}$ and wavelength $\protect\lambda _{\mathrm{L}}=1541.5\unit{nm}$, is coupled using a 5:95 optical coupler, which is connected between the points 'A' and 'B' shown in the inset of Fig. \ref{FigTempSU} of the main text. The temperature is $T=2.9 \unit{K}$, and the diode current is $%
		I_{\mathrm{D}}=150\unit{mA}$.}
	\label{FigHB}
\end{figure*}

\section{Short-time scale dynamics}
\label{SecSTSD}

The optical complex amplitude at time $t$ is expressed as $f\left( f_{%
\mathrm{L}}t\right) $, where $f_{\mathrm{L}}$ is the fiber loop frequency.
The function $f\left( x\right) $ can be Fourier expanded as%
\begin{equation}
f\left( x\right) =\sum_{n=-\infty }^{\infty }c_{n}e^{2\pi inx}\;,
\label{f(x)}
\end{equation}%
where $c_{n}$, which is expressed as $c_{n}=c_{n}^{\prime }+ic_{n}^{\prime
\prime }$ , where both $c_{n}^{\prime }$ and $c_{n}^{\prime \prime }$ are
real, is the complex amplitude of the $n$'th mode. The $L_{p}$ norm of $f$
is given by%
\begin{equation}
\left\Vert f\right\Vert _{p}=\left( \int_{0}^{1}\mathrm{d}x\;\left\vert
f\left( x\right) \right\vert ^{p}\right) ^{1/p}\;,  \label{Lp}
\end{equation}%
and the following holds%
\begin{equation}
\left\Vert f\right\Vert _{2}^{2}=\sum_{n=-\infty }^{\infty }\left\vert
c_{n}\right\vert ^{2}\equiv I\;,  \label{f L2}
\end{equation}%
where $I$\ is the total intensity, and%
\begin{equation}
\left\Vert f\right\Vert _{4}^{4}=\sum_{n^{\prime }-n^{\prime \prime
}+n^{\prime \prime \prime }-n^{\prime \prime \prime \prime }=0}c_{n^{\prime
}}^{{}}c_{n^{\prime \prime }}^{\ast }c_{n^{\prime \prime \prime
}}^{{}}c_{n^{\prime \prime \prime \prime }}^{\ast }\;.  \label{f L4}
\end{equation}%
As will be discussed below, the term $\left\Vert f\right\Vert _{4}^{4}$
plays an important role in the system's short-time dynamics. Below an upper bound imposed upon $\left\Vert f\right\Vert _{4}^{4}$ is derived [see inequality (\ref{f_4^4 bound}) below].

When the expansion (\ref{f(x)}) contains a finite number of non-vanishing
terms, $f\left( x\right) $ can be expressed as%
\begin{equation}
f\left( x\right) =\sum_{n\in N_{f}}c_{n}e^{2\pi inx}\;,
\end{equation}%
where $N_{f}=\left\{ n\in 
%TCIMACRO{\U{2124} }%
%BeginExpansion
\mathbb{Z}
%EndExpansion
:c_{n}\neq 0\right\} $ ($%
%TCIMACRO{\U{2124} }%
%BeginExpansion
\mathbb{Z}
%EndExpansion
$ denotes the set of integers), and $\left\Vert f\right\Vert _{4}^{4}$ as
[see Eq. (\ref{f L4})]%
\begin{eqnarray*}
\left\Vert f\right\Vert _{4}^{4} &=&\int_{0}^{1}\mathrm{d}x\;\left(
\sum_{n^{\prime },n^{\prime \prime }\in N_{f}}c_{n^{\prime }}c_{n^{\prime
\prime }}^{\ast }e^{2\pi i\left( n^{\prime }-n^{\prime \prime }\right)
x}\right) ^{2} \\
&=&\int_{0}^{1}\mathrm{d}x\;\left( \sum_{d}C_{d}e^{2\pi idx}\right) ^{2}\;,
\\
&&
\end{eqnarray*}%
where%
\begin{equation}
C_{d}=\sum_{\substack{ n^{\prime },n^{\prime \prime }\in N_{f}  \\ n^{\prime
}-n^{\prime \prime }=d}}c_{n^{\prime }}^{{}}c_{n^{\prime \prime }}^{\ast }\;.
\label{C_m}
\end{equation}%
Note that $C_{0}=\sum_{n^{\prime }}\left\vert c_{n^{\prime }}\right\vert
^{2}=\left\Vert f\right\Vert _{2}^{2}=I$, $C_{-d}^{{}}=C_{d}^{\ast }$ and%
\begin{eqnarray}
\left\Vert f\right\Vert _{4}^{4} &=&\sum_{d^{\prime },d^{\prime \prime
}}C_{d^{\prime }}^{{}}C_{d^{\prime \prime }}^{\ast }\int_{0}^{1}\mathrm{d}%
x\;e^{2\pi i\left( d^{\prime }-d^{\prime \prime }\right) x}  \notag \\
&=&\sum_{d}\left\vert C_{d}\right\vert ^{2}\;.  \notag \\
&&  \label{f_4^4}
\end{eqnarray}%
With the help of the Cauchy's Inequality one finds that [see Eq. (\ref{C_m})]%
\begin{equation}
\left\vert C_{d}\right\vert ^{2}\leq \mathcal{N}_{d}\sum_{\substack{ %
n^{\prime },n^{\prime \prime }\in N_{f}  \\ n^{\prime }-n^{\prime \prime }=d
}}\left\vert c_{n^{\prime }}^{{}}c_{n^{\prime \prime }}^{\ast }\right\vert
^{2}\;,
\end{equation}%
where $\mathcal{N}_{d}=\left\vert \left\{ \left( n^{\prime },n^{\prime
\prime }\right) \in N_{f}\times N_{f}:n^{\prime }-n^{\prime \prime
}=d\right\} \right\vert $ ($\left\vert S\right\vert $ denotes cardinality,
i.e. number of members, of a given set $S$), and thus [see Eqs. (\ref{f L2})
and (\ref{f_4^4})]%
\begin{eqnarray}
\left\Vert f\right\Vert _{4}^{4} &=&\left\vert C_{0}\right\vert
^{2}+\sum_{d^{\prime }\neq 0}\left\vert C_{d^{\prime }}\right\vert ^{2} 
\notag \\
&\leq &\left( 1+\max_{d\neq 0}\mathcal{N}_{d}\right) \left\Vert f\right\Vert
_{2}^{4}\;.  \notag \\
&&  \label{f_4^4 bound}
\end{eqnarray}

Consider the case where the term proportional to $g_{m}$ in Eq. (\ref{a_m
dot}) in the main text can be disregarded. For this case, and when noise is
taken into account, the set of coupled equations can be expressed as [see
Eq. (\ref{a_m dot}) in the main text]%
\begin{equation}
\dot{c}_{m}=-\partial _{m}^{\ast }\mathcal{H}+\xi _{m}\;,  \label{c_m dot H}
\end{equation}%
where $\partial _{m}$, which is given by%
\begin{equation}
\partial _{m}=\frac{\partial }{\partial c_{m}}=\frac{1}{2}\left( \frac{%
\partial }{\partial c_{m}^{\prime }}-i\frac{\partial }{\partial
c_{m}^{\prime \prime }}\right) \;,
\end{equation}%
is the Wirtinger derivative (note that $\partial _{m}c_{m}^{{}}=1$ and $%
\partial _{m}c_{m}^{\ast }=0$), the Hamiltonian is given by $\mathcal{H}=%
\mathcal{H}_{\mathrm{d}}+\mathcal{H}_{\mathrm{c}}$, where the dispersion
Hamiltonian $\mathcal{H}_{\mathrm{d}}$ is given by $\mathcal{H}_{\mathrm{d}%
}=-\Gamma _{\mathrm{d}}\sum_{n^{\prime }}n^{\prime 2}c_{n^{\prime
}}^{{}}c_{n^{\prime }}^{\ast }$, and the intermode coupling Hamiltonian $%
\mathcal{H}_{\mathrm{c}}$ is given by [see Eq. (\ref{f L4})]%
\begin{equation}
\mathcal{H}_{\mathrm{c}}=\frac{\Gamma _{\mathrm{c}}}{2}\left\Vert
f\right\Vert _{4}^{4}\;.  \label{H_c}
\end{equation}%
The complex white noise terms $\xi _{m}=\xi _{m}^{\prime }+i\xi _{m}^{\prime
\prime }$, where both $\xi _{m}^{\prime }$ and $\xi _{m}^{\prime \prime }$
are real, satisfy the relations $\left\langle \xi _{m}^{\prime }\left(
t\right) \xi _{m}^{\prime }\left( t^{\prime }\right) \right\rangle
=\left\langle \xi _{m}^{\prime \prime }\left( t\right) \xi _{m}^{\prime
\prime }\left( t^{\prime }\right) \right\rangle =2\tau \delta \left(
t-t^{\prime }\right) $ and $\left\langle \xi _{m^{\prime }}^{\prime }\left(
t\right) \xi _{m^{\prime \prime }}^{\prime \prime }\left( t^{\prime }\right)
\right\rangle =0$, where $\tau \ $is positive, and where angle brackets
denote time averaging. The steady state solution of Eq. (\ref{c_m dot H}) is
given by Eq. (7.251) of Ref. \cite{S_Buks_SPLN}.

The above finding (\ref{H_c}) that $\mathcal{H}_{\mathrm{c}}$ is
proportional to $\left\Vert f\right\Vert _{4}^{4}$, together with the upper bound given by inequality (\ref{f_4^4 bound}), suggest that the intermode
coupling contribution $\mathcal{H}_{\mathrm{c}}$ to the total Hamiltonian
can be minimized (for a given total intensity $\left\Vert f\right\Vert
_{2}^{2}=I$) by selecting the set of excited modes $N_{f}$ such that $%
\max_{d\neq 0}\mathcal{N}_{d}$ is minimized. In particular, the case $%
\max_{d\neq 0}\mathcal{N}_{d}=1$ is discussed below.

When the condition $\max_{d\neq 0}\mathcal{N}_{d}=1$ is satisfied the set $%
N_{f}$ is said to be SU. Alternatively, the term SU can be defined as
follows. Consider the equation%
\begin{equation}
n^{\prime }-n^{\prime \prime }=n^{\prime \prime \prime }-n^{\prime \prime
\prime \prime }\ ,  \label{n pair}
\end{equation}%
where $n^{\prime },n^{\prime \prime },n^{\prime \prime \prime },n^{\prime
\prime \prime \prime }\in N_{f}$. When Eq. (\ref{n pair}) is unsolvable,
unless $n^{\prime }=n^{\prime \prime \prime }$ and $n^{\prime \prime
}=n^{\prime \prime \prime \prime }$, the set $N_{f}$ is said to be SU. For
this case all spacings between pairs of elements belonging to $N_{f}$ are
unique, i.e. $\max_{d\neq 0}\mathcal{N}_{d}=1$, and the upper bound given by
inequality (\ref{f_4^4 bound}) yields $\mathcal{H}_{\mathrm{c}}\leq \Gamma _{%
\mathrm{c}}I^{2}$.

\section{The measured USOC wavelength sequence $\left\{ \protect\lambda %
_{k}\right\} $}
\label{SecSequence}

The measured USOC wavelength sequence $\left\{ \lambda _{k}\right\} $ is
presented by Fig. \ref{FigComb_in_In} in the main text. The red solid line
shown in Fig. \ref{FigComb_in_In}(b) in the main text is calculated using
the relation for $k\geq 1$%
\begin{equation}
n_{k}=\nu \log p_{k}\ ,  \label{i_n}
\end{equation}%
where $p_{k}$ is the $k$'th prime number. For this measurement, the
dimensionless pre-factor $\nu $ is found by fitting to be given by $\nu
=5721 $. The comparison between the measured values of $i_{k}=\left(
f_{0}-f_{k}\right) /f_{\mathrm{L}}$ and the calculated values of $n_{k}$
obtained from Eq. (\ref{i_n}) yields a good agreement [see Fig. \ref%
{FigComb_in_In}(b) in the main text]. The level of agreement is quantified
by the parameter $\varepsilon =k_{\mathrm{m}}^{-1}\sum_{k=1}^{k_{\mathrm{m}%
}}\left\vert \left( i_{k}-n_{k}\right) /i_{k_{\mathrm{m}}}\right\vert $,
where $k_{\mathrm{m}}$ is the number of peaks that can be reliably resolved.
For the data shown in Fig. \ref{FigComb_in_In}) in the main text $k_{\mathrm{%
m}}=150$ and $\varepsilon =0.004$. The USOC shown in Fig. \ref{FigComb_in_In}
in the main text has been obtained in a slow cooling process with a fixed
applied diode current $I_{\mathrm{D}}$. Note that larger deviation between
data and Eq. \ref{i_n} is observed in other cases. One example is a USOC
that is obtained by switching off, and then abruptly switching on, the diode
current $I_{\mathrm{D}}$, while keeping the temperature at its base value.
For this case typically $\varepsilon \simeq 0.01$.

\section{RU and SU sets}
\label{SecRUSU}

The above-discussed experimental finding that $i_{k}=\left(
f_{0}-f_{k}\right) /f_{\mathrm{L}}\simeq n_{k}$, where $n_{k}$ is given by
Eq. (\ref{i_n}), implies that the set $N_{f}$ of excited mode indices
corresponding to the measured USOC is nearly SU [recall the fundamental
theorem of arithmetic, and that $\log x+\log y=\log \left( xy\right) $]. As
was shown above, this observation suggests that the USOC wavelength sequence
is selected in a way that minimizes the energy associated with intermode
coupling (for a given total intensity $I$). On the other hand, other USOC
wavelength sequences can give rise to other SU sets $N_{f}$. Below we show
that for a given width of the lasing band, the number of peaks of the
experimentally observed USOC has the same order of magnitude as the largest
possible number. The analysis below also addresses the question what
determines the value of the dimensionless coefficient $\nu $ in Eq. (\ref%
{i_n}).

A set $A=\left\{ e_{1},e_{2},\cdots ,e_{N}\right\} $ is said to be RU if all
ratios between pairs of elements belonging to $A$ are unique, i.e. if the
equation $e_{n^{\prime }}/e_{n^{\prime \prime }}=e_{n^{\prime \prime \prime
}}/e_{n^{\prime \prime \prime \prime }}$ is unsolvable, unless $n^{\prime
}=n^{\prime \prime \prime }$ and $n^{\prime \prime }=n^{\prime \prime \prime
\prime }$. Similarly, as was already defined above, a set $A=\left\{
e_{1},e_{2},\cdots ,e_{N}\right\} $ is said to be SU if all spacings between
pairs of elements belonging to $A$ are unique, i.e. if the equation $%
e_{n^{\prime }}-e_{n^{\prime \prime }}=e_{n^{\prime \prime \prime
}}-e_{n^{\prime \prime \prime \prime }}$ is unsolvable, unless $n^{\prime
}=n^{\prime \prime \prime }$ and $n^{\prime \prime }=n^{\prime \prime \prime
\prime }$.

The set $A_{\mathrm{p}}=\left\{ p_{1},p_{2},p_{3},p_{4},\cdots ,p_{N}\right\}
$ of the first $N$ prime numbers is RU (the $n$'th prime number is denoted
by $p_{n}$). The set $A_{\mathrm{p}}$ can be used for the generation of an
SU set $A_{\mathrm{g}}=\left\{ \log p_{1},\log p_{2},\cdots ,\log
p_{N}\right\} $. The set of real numbers $A_{\mathrm{g}}$ can be mapped into
a set of integers $A_{ \mathrm{gi}}=\left\{ a_{1},a_{2},\cdots
,a_{N}\right\} $, where%
\begin{equation}
a_{n}=\left\lfloor \nu \log p_{n}+1/2\right\rfloor \ ,  \label{a_n SU}
\end{equation}%
and where $\nu >0$ is a constant ($\left\lfloor x+1/2\right\rfloor $ is the
nearest integer to $x$, where $\left\lfloor x\right\rfloor $ is the floor of 
$x$).

\begin{figure}[tbp]
\begin{center}
\includegraphics[width=3.2in,keepaspectratio]{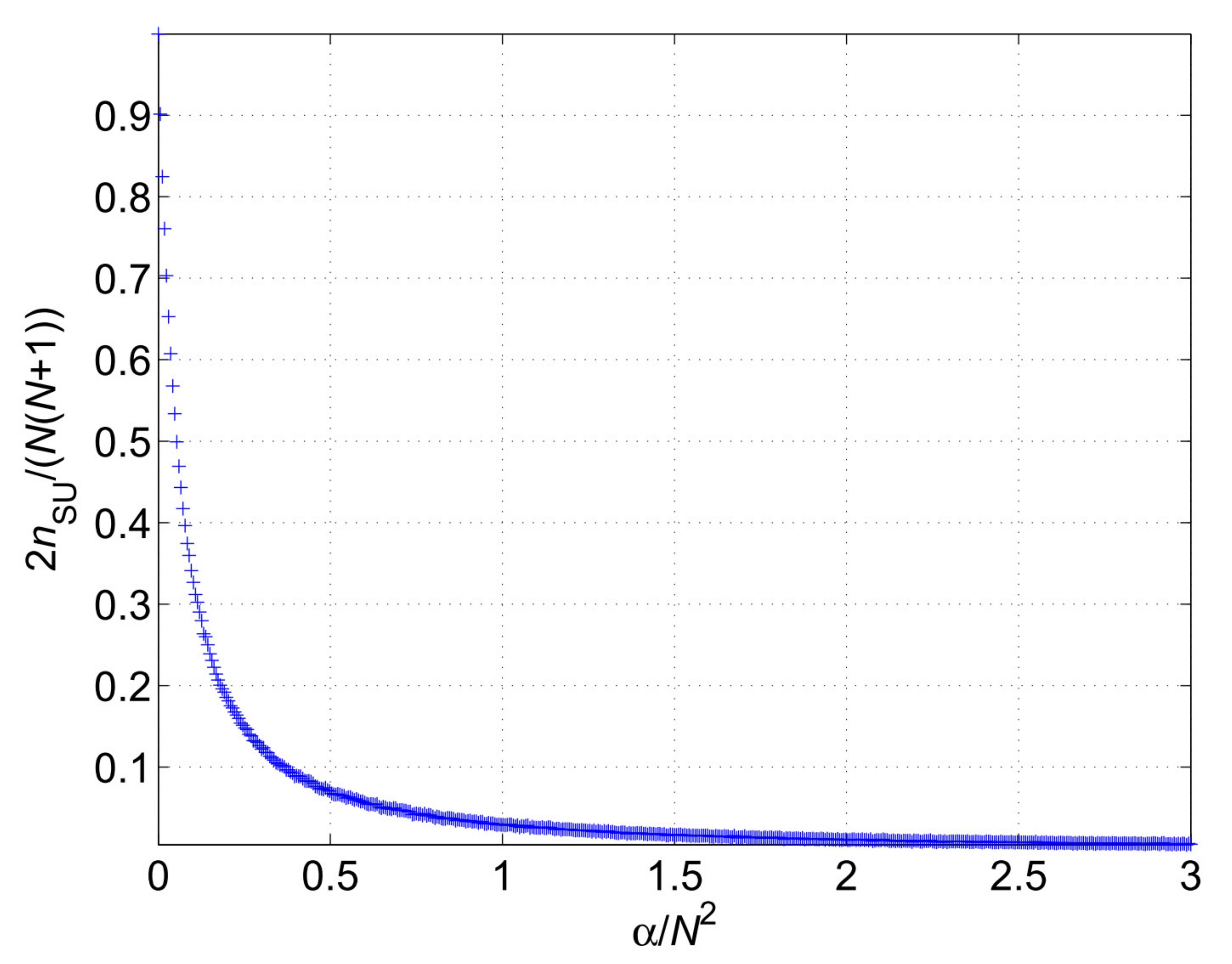}
\end{center}
\caption{{}A plot of $2n_{\mathrm{SU}}/\left( N\left( N+1\right) \right) $
as a function of $\protect\nu /N^{2}$ for the case $N=1500.$}
\label{FignSU}
\end{figure}

For what values of $\nu $ the set $A_{\mathrm{gi}}$ becomes SU? Let $S_{%
\mathrm{gi}}=\left\{ a_{n^{\prime }}-a_{n^{\prime \prime }}|N\geq n^{\prime
}>n^{\prime \prime }\geq 1\right\} $ be the set of positive spacings between
pairs of elements belonging to $A_{\mathrm{gi}}$. The integer $n_{\mathrm{SU}%
}$ is defined by%
\begin{equation}
n_{\mathrm{SU}}=\frac{N\left( N+1\right) }{2}-\left\vert S_{\mathrm{gi}%
}\right\vert \ .
\end{equation}%
When $n_{\mathrm{SU}}=0$ the set $A_{\mathrm{gi}}$ is SU. A plot of $2n_{%
\mathrm{SU}}/\left( N\left( N+1\right) \right) $ as a function of $\nu
/N^{2} $ is shown in Fig. \ref{FignSU}. The plot in Fig. \ref{FignSU}
suggests that for $\nu \gtrsim N^{2}$ the set $A_{\mathrm{gi}}$ becomes SU.
The relation $\nu =N^{2}$ together with the number of observed USOC peaks $%
\simeq 150$ yields the value of $22500$ for $\nu $. This rough estimate is
about $3.9$ times the value of $\nu =5721$ that is extracted from the fit
between data and Eq. \ref{i_n} (see Fig. \ref{FigComb_in_In} in the main
text).

As was shown above, the experimentally observed USOC yields the SU set $A_{%
\mathrm{gi}}$ given by Eq. (\ref{a_n SU}) [compare with Eq. \ref{i_n}]. For
the general case, a comb can be characterized by the number of peaks within
the lasing bandwidth, which depends on the packing factor $\chi $ of the
corresponding SU set of integers. Below we define the packing factor $\chi $%
, derive an upper bound for $\chi $, and compare the bound with the packing
factor $\chi \left( A_{\mathrm{gi}}\right) $ of the SU set $A_{\mathrm{gi}}$%
. As is shown below, $\chi \left( A_{\mathrm{gi}}\right) $ has the same
order of magnitude as the largest possible value of $\chi $ of any SU set
having the same number $N$ of elements.

The integer $a_{n}\in A_{\mathrm{gi}}$ (\ref{a_n SU}) can be expressed as%
\begin{equation}
a_{n}=\nu \log p_{n}+\epsilon _{n}\ ,  \label{a_n epsilon SU}
\end{equation}%
where $\left\vert \epsilon _{n}\right\vert \leq 1/2$. The equation $%
a_{n^{\prime }}-a_{n^{\prime \prime }}=a_{n^{\prime \prime \prime
}}-a_{n^{\prime \prime \prime \prime }}$ can be rewritten as%
\begin{equation}
0=\nu \log \varrho +\varepsilon \ ,  \label{C SU}
\end{equation}%
where $\varrho =\left( p_{n^{\prime }}/p_{n^{\prime \prime }}\right) /\left(
p_{n^{\prime \prime \prime }}/p_{n^{\prime \prime \prime \prime }}\right) $,
and where $\varepsilon $, which is given by $\varepsilon =\epsilon
_{n^{\prime }}-\epsilon _{n^{\prime \prime }}-\epsilon _{n^{\prime \prime
\prime }}+\epsilon _{n^{\prime \prime \prime \prime }}$, is bounded by $%
\left\vert \varepsilon \right\vert \leq 2$. Consider a nontrivial solution
of Eq. (\ref{C SU}), i.e. a solution for which $\varrho \neq 1$. Without
loss of generality, it is assumed that $\varrho >1$, i.e. $p_{n^{\prime
}}p_{n^{\prime \prime \prime }}>p_{n^{\prime \prime }}p_{n^{\prime \prime
\prime \prime }}$. Since $p_{n}$ are all integers, this implies that $%
p_{n^{\prime }}p_{n^{\prime \prime \prime }}-p_{n^{\prime \prime
}}p_{n^{\prime \prime \prime \prime }}\geq 1$. By using the relation $%
\varrho -1=\left( p_{n^{\prime }}p_{n^{\prime \prime \prime \prime
}}-p_{n^{\prime \prime }}p_{n^{\prime \prime \prime }}\right) /\left(
p_{n^{\prime \prime }}p_{n^{\prime \prime \prime }}\right) $ one finds that $%
\varrho =1+\varrho -1\geq 1+1/\left( p_{n^{\prime \prime }}p_{n^{\prime
\prime \prime }}\right) \geq 1+p_{N}^{-2}$. Thus $\left\vert \log \varrho
\right\vert \geq p_{N}^{-2}$ (it is assumed that $N\gg 1$), hence $A_{ 
\mathrm{gi}}$ is SU provided that $\nu \geq 2p_{N}^{2}$.

According to the prime number theorem, for $n\geq 6$ the following holds 
\cite{S_Dusart_411}%
\begin{equation}
\zeta _{n}-\frac{1}{2}<\frac{p_{n}}{n}<\zeta _{n}+\frac{1}{2}\ ,
\label{PNT zeta}
\end{equation}%
where $\zeta _{n}=\log n+\log \log n-1/2$. Using the approximation [see
inequality (\ref{PNT zeta})]%
\begin{equation}
p_{n}\simeq n\zeta _{n}\ ,  \label{p_n zeta}
\end{equation}%
one finds for the case $\nu =2p_{N}^{2}$ that%
\begin{equation}
a_{N}\simeq 2N^{2}\zeta _{N}^{2}\log \left( N\zeta _{N}\right) \ .
\label{a_N PNT}
\end{equation}

For a general SU set of integers $A=\left\{ a_{1},a_{2},\cdots
,a_{N}\right\} $, where $1\leq a_{1}<a_{2}<\cdots <a_{N}$, the packing
factor $\chi \left( A\right) $ is defined by $\chi \left( A\right) =N/a_{N}$%
. An upper bound upon $\chi \left( A\right) $ is derived below. For a given
positive integer $d$, let $S_{d}$ be the number of pairs $\left(
a_{n^{\prime }},a_{n^{\prime \prime }}\right) \in A^{2}$ such that $%
a_{n^{\prime \prime }}-a_{n^{\prime }}=d$. Since $A$ is SU, $S_{d}\in
\left\{ 0,1\right\} $ for any positive integer $d$. The number of distinct
positive integers $d$, such that $S_{d}=1$ is $N\left( N-1\right) /2$ (i.e.
the number of ordered pairs of distinct element in $A$). On the other hand, $%
a_{n^{\prime \prime }}-a_{n^{\prime }}\leq a_{N}-1$ for any $a_{n^{\prime
}},a_{n^{\prime \prime }}\in A$, and thus $N\left( N-1\right) /2\leq a_{N}-1$%
. This condition imposes an upper bound upon the packing factor%
\begin{equation}
\chi \left( A\right) \leq \frac{1}{\frac{N}{2}-\frac{1}{2}+N^{-1}}\ .
\label{chi B}
\end{equation}%
This bound can be compared with the packing factor of the SU set $A_{\mathrm{%
gi}}$, which for the case $\nu =2p_{N}^{2}$ is given by [see Eq. (\ref{a_N
PNT})]%
\begin{equation}
\chi \left( A_{\mathrm{gi}}\right) \simeq \frac{1}{2N\zeta _{N}^{2}\log
\left( N\zeta _{N}\right) }\ .  \label{chi gi}
\end{equation}%
Hence, to leading order $\chi =O\left( N^{-1}\right) $ for both the upper
bound (\ref{chi B}) and for the set $A_{\mathrm{\ gi}}$ (\ref{chi gi}). In
other words, for large $N$ the packing factor of the SU set $A_{\mathrm{\ gi}%
}$ has the same order of magnitude as the largest possible packing factor.

\end{document}